# In-flight performances of the BeppoSAX Gamma-Ray Burst Monitor

M. Feroci[a,1], F. Frontera[b,c], E. Costa[a], D. Dal Fiume[c],
L. Amati[a], L. Bruca[d], M.N. Cinti[a], A. Coletta[d], P. Collina[b], C. Guidorzi[b],
L. Nicastro[c], M. Orlandini[c], E. Palazzi[c], M. Rapisarda[a,e], G. Zavattini[b] and R.C. Butler[f]

[a]Istituto di Astrofisica Spaziale - C.N.R. - Frascati, Italy
[b]Dipartimento di Fisica - Universita' di Ferrara - Italy
[c]Istituto Tecnologie e Studio Radiazioni Extraterrestri - C.N.R. - Bologna, Italy
[d]BeppoSAX Scientific Operation Center - Nuova Telespazio - Roma, Italy
[e]Divisione di Neutronica - ENEA - Frascati, Italy
[f]Agenzia Spaziale Italiana - Roma, Italy

**ABSTRACT**

The Italian-Dutch satellite for X-ray Astronomy BeppoSAX is successfully operating on a 600 km equatorial orbit since May 1996. We present here the in-flight performances of the Gamma Ray Burst Monitor experiment during its first year of operation. The GRBM is the secondary function of the four CsI(Na) slabs primarily operating as an active anticoincidence of the PDS hard X-ray experiment. It has a geometric area of about 4000 cm$^2$ but, due to its location in the core of the satellite its effective area is dependent on the energy and direction of the impinging photons. A dedicated electronics allows to trigger on cosmic gamma-ray bursts. When the trigger condition is satisfied the light curve of the event is recorded from 8 s before to 98 s after the trigger time, with a maximum time resolution of 0.48 ms, in an energy band of 40-700 keV. As an instrument housekeeping the 1 s event ratemeter of the same detectors in the same energy band is stored regardless the trigger condition, allowing for an off-line detection of non-triggered events. Finally, the onboard software collects the event count rate that is used as anticoincidence, i.e. the events above a given energy threshold, typically kept at 100 keV. The flight-data screening is in progress, in order to extract real Gamma Ray Bursts from the many sources of background. Already many results have been obtained, as those GRBs detected simultaneously with the Wide Field Cameras onboard BeppoSAX itself.

## 1. INTRODUCTION

The BeppoSAX[1] satellite has been successfully launched by the Italian Space Agency from the Cape Canaveral launch base on April 30, 1996. The two-stages Atlas-Centaur rocket has put the satellite on a 4° inclination, 600 km altitude orbit. The scientific payload of the satellite is composed by four narrow field X-ray telescopes[2,3,4,5] (NFI), covering the energy range from 0.1 to 300 keV. Their common optical axis forms a 90° angle with the optical axis of the two wide field experiments, the two coded mask Wide Field Cameras[6] (WFCs, 40°x40° zero response field of view, 2-26 keV, watching 180° one from each other) and the Gamma-Ray Burst Monitor[7] (GRBM, open field of view, 40-700 keV), watching a portion of the sky that includes the FOV of the WFCs. The satellite pointing strategy is usually such that the NFI are pointed to some celestial source, while the WFC and GRBM are operated in slave mode. Given the better capabilities of the WFCs for galactic sources rather than for extragalactic targets, the pointing program is optimized in order to point at least one of the two WFCs to the Galactic Plane, where most of the galactic X-ray sources are located.

[1] Correspondence: M.F., Istituto di Astrofisica Spaziale - C.N.R. - Via Enrico Fermi 21 - C.P. 67 - I-00044 Frascati, Italy.
Fax: +39-6-9416847, Voice: +39-6-9424589, E-mail: feroci@saturn.ias.fra.cnr.it

The simultaneous presence of the coaligned WFCs and GRBM, together with the NFI onboard the same satellite has revealed to be the breakthrough in the gamma-ray bursts (GRBs) astronomy. The GRBs are intense, short, unpredictable flashes of gamma-rays with typical energies of the order of few hundreds keV[8]. They are still unidentified objects with a distribution in the sky that appears to be isotropic, but not homogeneous due to a paucity of the weaker events. After the pioneeristic work carried out by the experiments onboard the Vela[9] satellites, the first systematic observation of GRBs was performed by the Konus experiment[10] onboard Venera 11 and Venera 12 spacecrafts, and at a major extent by the Burst And Transient Sources Experiment[11] (BATSE) onboard the US Compton Gamma Ray Observatory (CGRO). Its GRB catalog[12] presently enumerates about 2,000 GRBs, with fluences ranging from $10^{-8}$ to $10^{-4}$ ergs cm$^{-2}$.

The GRBM is able to detect GRBs with a peak flux as weak as $10^{-8}$ ergs cm$^{-2}$ s$^{-1}$ and trigger those above $10^{-7}$ with its onboard selection logic. If the GRB is in the FOV of one of the two WFCs, then it can be promptly localized with an accuracy as high as 3 arcminutes error radius by detecting its X-ray emission. Even if the satellite is visible from ground only once per orbit (97 minutes), with such an accurate position BeppoSAX can be reoriented to observe the GRB location as shortly as 6 hours after the GRB detection. This procedure has lead to one of the most important discoveries of the GRB astronomy, the detection of the X-ray[13] and optical[14] counterparts of GRB970228. Three more GRBs have been promptly pointed by BeppoSAX narrow field X-ray telescopes: GRB970111[15] after 16 hours has revealed no X-ray counterpart; GRB970402[16] has shown a X-ray counterpart very similar to GRB970228, but no optical counterpart has been detected; both X-ray[17] and optical[18] counterparts have been detected for GRB970508, but with a time behaviour rather different from the others. In the latter case the measurement of its optical spectrum[19] has allowed to place the optical source to a minimum redshift of 0.835, that is outside our own Galaxy, solving one of the greatest astronomical mysteries of the last two decades.

In this paper we present the in-flight performances of the instrument that is the first link of this chain, the GRBM. It is a not dedicated experiment, that has surprisingly revealed astrophysical capabilities comparable to the dedicated experiment BATSE. It detects tens of GRBs every month, confirming GRB events detected by BATSE and other experiments like those onboard Ulysses[20] and Wind[21] satellites.

## 2. THE INSTRUMENT

The GRBM is an experiment derived from a secondary function of the anticoincidence (AC) detectors of the high energy experiment Phoswich Detection System[5] (PDS, 15-300 keV). This is composed of four identical, coaligned NaI/CsI scintillators in a phoswich configuration, with a 1.3° collimated field of view. In Figure 1 a sketch of the PDS/GRBM is presented together with the others BeppoSAX experiments. The AC is used to reject those background events that give a signal both in one of the four phoswich units and in one of the AC detectors. The latter are four CsI(Na) shields forming a square box, surrounding the main PDS detectors. Each AC detector is 1 cm thick and has dimensions 27.5 x 41.3 cm, for a geometric area of about 1136 cm$^2$ and an open field of view. The four CsI detectors composing the GRBM are optically independent. Each one is composed by two optically coupled halves, whose light is seen by two independent photomultipliers Hamamatsu. A calibration source for each of the four detectors is provided. It is a light source obtained with an Am$^{241}$ source embedded in a NaI crystal. Its light is then seen as a permanent source in the detector through a grey filter.

When it was at the design level, the large area of the AC detectors suggested their use as GRB monitor. A proper electronics was therefore realized. It consists of a dedicated ADC converter to which the multiplexed signals coming from each of the lateral shields are fed, after a set of analog thresholds has been satisfied. In particular, for each of the four lateral shields a Low Level Threshold (LLT) can be adjusted by telecommand among 16 steps, nominally ranging from 20 to 90 keV, and an Upper Level Threshold (ULT) can be adjusted in 8 steps, from nominal 200 to 600 keV (actually, as experimented during the on-ground calibrations, the ADC records event energies up to 700 keV). The signals detected in each shield between the LLT and the ULT are continuously counted, with a time resolution of 1 s, and recorded by a proper instrument housekeeping ("GRBM"). Another similar housekeeping counter ("LS") records, separately for each shield, the count rate detected above an AC Threshold (ACT), that is used as a lower level discriminator for any AC signal, adjustable on 8 steps between nominal 100 and 300 keV.

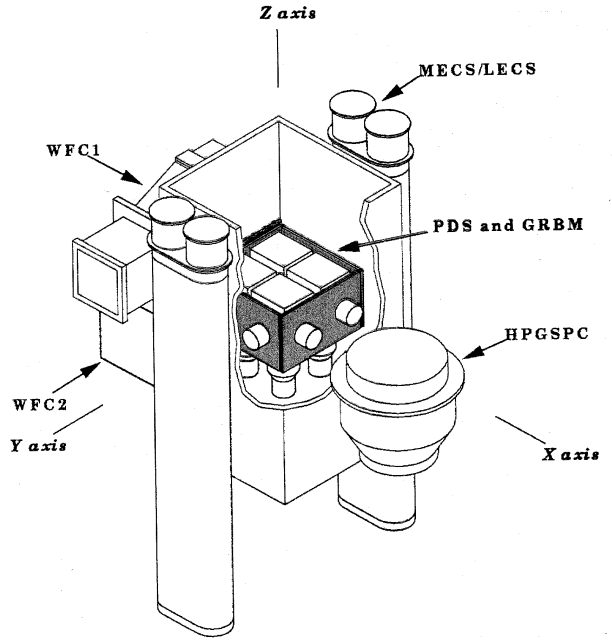

**Figure 1:** The optical bench of BeppoSAX with the four NFI instruments, the GRBM and the two WFCs.

Due to the nature of GRBs an onboard specific trigger criterion is needed. The GRBM trigger operates on the signals detected between the LLT and ULT. With a time resolution on 7.8125 ms a moving average is continuously computed on a Long Integration Time (LIT) that is adjustable between 8 and 128 s. The counts in a Short Integration Time (SIT, adjustable between 7.8125 ms and 4 s) are compared to the moving average, and if they exceed a $n\sigma$ level (where $\sigma$ is the Poissonian standard deviation and $n$ can be 4, 8 or 16) then the trigger condition is satisfied for that shield. If the same condition is simultaneously active for at least two shields, then the GRBM trigger condition is satisfied and the following time profiles (High Resolution Time Profiles) are stored, separately for each of the four shields:
 a) 8 s before the trigger time, with a time resolution of 7.8125 ms
 b) 10 s from the trigger time, with a time resolution of 0.48828125 ms
 c) 88 s starting from 10 s after the trigger time, with a time resolution of 7.8125 ms

Independently of the trigger condition, besides the GRBM and LS housekeepings, 256-channel energy spectra are stored onboard every 128 s, separately for each of the four shields. These are integrated over fixed time intervals, and are therefore mainly useful for calibration purposes. In fact, given the typical time scales of cosmic GRBs the housekeeping spectral data can only be used for obtaining the average energy spectrum of bright GRBs, when the GRB fluence is visible with respect to the 128 s integrated background. However the presence of the two housekeeping ratemeters, GRBM and LS, allows for some spectral reconstruction. The GRBM and LS ratemeters are overlapped in the nominal energy band from 100 to 700 keV. Therefore, if one assumes[22] to be able to correct for the counts above 700 keV, then one can use the difference between GRBM and LS in order to extract a 2-channel spectrum, with a time resolution of 1 s. It is worth noticing that the events that are recorded in these two ratemeters, are exactly the same in the overlapped energy range, and are therefore completely covariant, significantly lowering the statistical error on their difference.

## 3. IN-FLIGHT CALIBRATION AND PERFORMANCES

The in-flight operations of the BeppoSAX GRBM were started on May 29, 1996 as part of the Commissioning Phase of the PDS instrument. The GRBM was very cautiously switched on for the first time during the satellite visibility (passage over the BeppoSAX Ground Station in Malindi) #411. The high voltage power supplies of the photomultipliers were set, one at time, at their lowest value in order to check the proper functioning of each one. After this successful test, the high voltages were turned on at their nominal value, and as a final step all of them were switched on together at their standard operation value with the whole PDS instrument during the visibility #427 on May 30, 1997, 18:50 UT. The in-flight performances of the GRBM experiment during its Commissioning Phase were nominal.

On July 1996 the Scientific Verification Phase of the BeppoSAX science instrumentation started. While the Commissioning Phase was dedicated to the functional tests, this phase was devoted to study the proper scientific operation of the instrumentation. For what concerns the GRBM the main task was to learn about the radiation environment in which the experiment had to operate and consequently identify the proper setting of all energy thresholds and trigger parameters. In Figure 2 we show a typical time profile along two satellite orbits of the four GRBM and LS ratemeters with a default set of thresholds. The very good stability and low modulation of the background along the orbit in the low energy ratemeters (GRBM) is apparent, as well as the number of spurious signals ("spikes") that make the time variation of the ratemeters not consistent with simple counting statistics. These spikes are mainly due to cosmic rays of high atomic number Z (or at the end of range) exciting metastable states in the CsI crystal, giving rise to a phosphorescence on short time scales, but long enough when compared to the response of the detector electronics, therefore allowing the electronics to detect a high number of counts. From the behaviour of the simultaneous LS ratemeters, however, we see that the typical amplitude of the signals composing the spikes corresponds to a small equivalent energy. This fact was confirmed by the in-flight thresholds calibration, that showed that they progressively diminish as the energy threshold rises.

Since the satellite is in a low-earth, almost equatorial orbit it crosses the South Atlantic Geomagnetic Anomaly (SAGA) once per orbit, at a latitude that varies during the day from about +4° to -4° because of the Earth rotation. A Particle Monitor (PM) has been therefore included in the PDS experiment in order to detect the instantaneous particle flux and to guard against sudden increases that may damage the experiment. It is thence also a very good monitor of the "proximity" of the satellite passage to the high particle density regions of the SAGA. In Figure 3 we show the time profile of the LS in time coincidence with the PM count rate profile for the duration of about 28 hours. From the behaviour of the PM counting rate it is evident the effect of the orbit precession during the day, and from the correlated LS counting rate it is also clear the effect of the CsI activation during the "deep" passages over the SAGA.

At the beginning of the mission the default values for the energy thresholds were 1 (nominally corresponding to 24.67 keV, and to a calibrated value 32.8 keV) for the LLT of the four shields and 6 (corresponding to a calibrated energy of 604 keV for the photopeak) for the ULT. The ACT was set as a default at 0 (lowest value, nominal 100 keV). The trigger condition at that time was set with LIT=128 s, SIT=4 s and $n$=8. Considering also that the small efficiency[22] of the GRBM detectors at the lower energies, due to the opacity of the materials located in their field of view, it was decided to rise the LLT to its step 3, that is a calibrated value of 42.5 keV, for the four shields. This new threshold setting is therefore almost ineffective from the point of view of the source detection, but it largely suppresses the noise due to spikes and also reduces the background count rate to a mean value of about 1000 counts s$^{-1}$. We note, however, that single spikes do not affect the trigger condition, that requires the time coincidence of the trigger on at least two shields, but with a SIT set at 4 s the probability of getting by chance a simultaneous trigger condition on two shields is enhanced by the fact that the temporal coincidence is extended to 4s. Furthermore, we note that the typical intensity of the spikes largely exceeds the 8σ level over the background, and therefore the $n$=8 condition cannot be considered an efficient discrimination method of the real GRB signals from the spikes.

Another important calibration therefore concerned the triggering parameters. With a SIT value of 4 s and $n$=8 the trigger condition selects GRB with high intensity and long duration (i.e. one second or more). Given the typical duration of GRBs[8] the SIT was lowered to 1s in order to be able to catch also shorter/weaker events. The high value of $n$ was no more needed after the GRBM count rate was reduced to a poissonian statistics with the suppression of most of the spikes. Taking into account that the trigger condition must be satisfied simultaneously on at least two shields in order to give a proper GRB trigger, then the value of 4 for $n$ can be considered safe enough against false triggers. In fact, considering that in one orbit (97 minutes) we get about 4000 s of good data (due to the data acquisition interruption at the passage over the South Atlantic Geomagnetic Anomaly, SAGA), following a Gaussian statistics we expect that the count rate exceeds by 4σ the average background count rate with a probability of 6x10$^{-5}$ each second in one shield. Since the non-source events must be uncorrelated on two shields the probability of having a GRB trigger by chance in Gaussian statistics is about 2x10$^{-8}$ each second, that is of the order of 8x10$^{-5}$ each orbit.

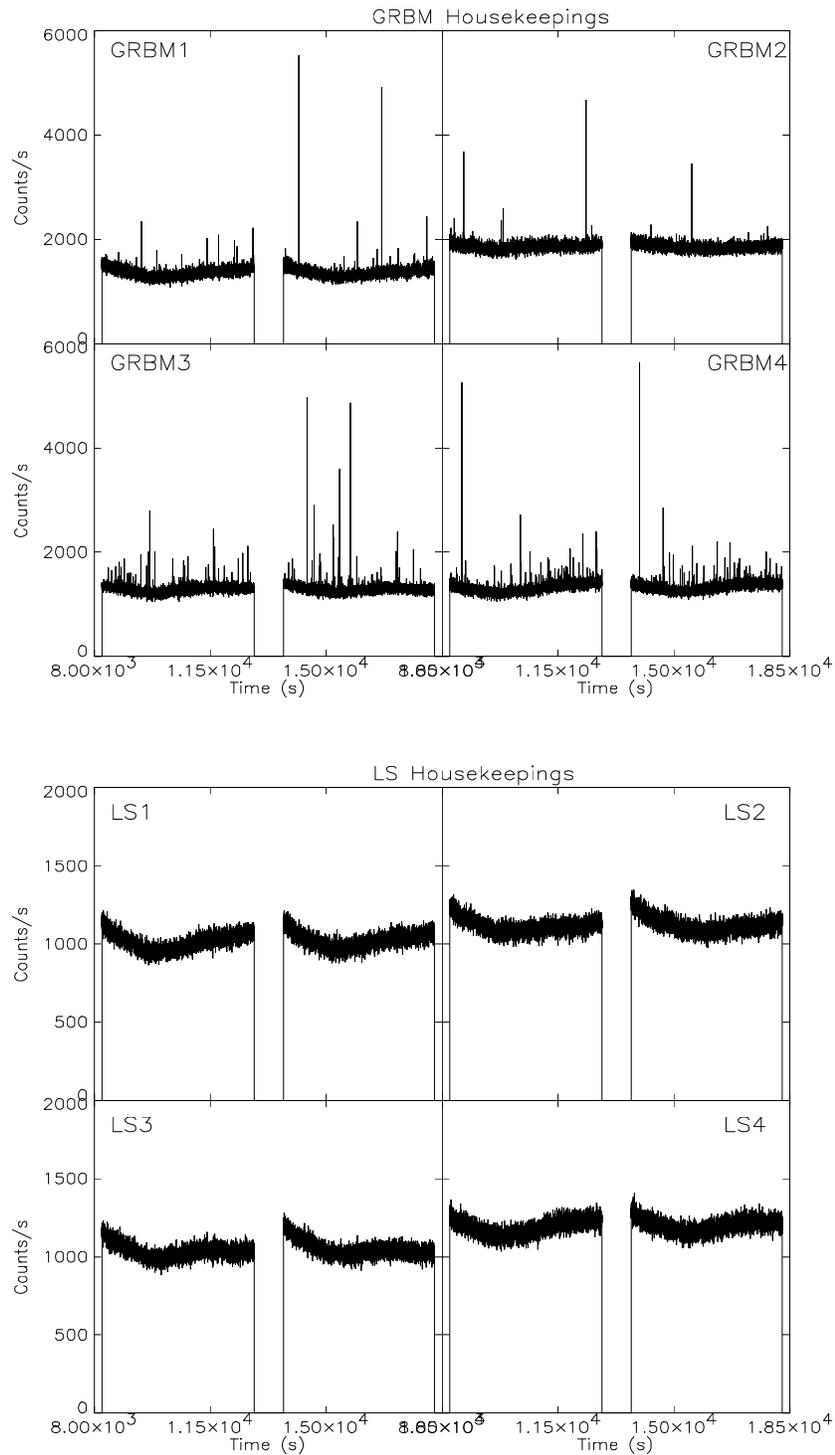

**Figure 2:** Typical on-orbit background level of the four GRBM and LS ratemeters of the GRBM experiment with GRBM energy thresholds LLT=1 and ULT=6, and ACT=0. The data gaps are due to the satellite passage close to the South Atlantic Anomaly and therefore identify the single orbits. The low energy channel (GRBM) shows the presence of several spikes, whose associated energy is within the energy difference between the GRBM LLT and the ACT.

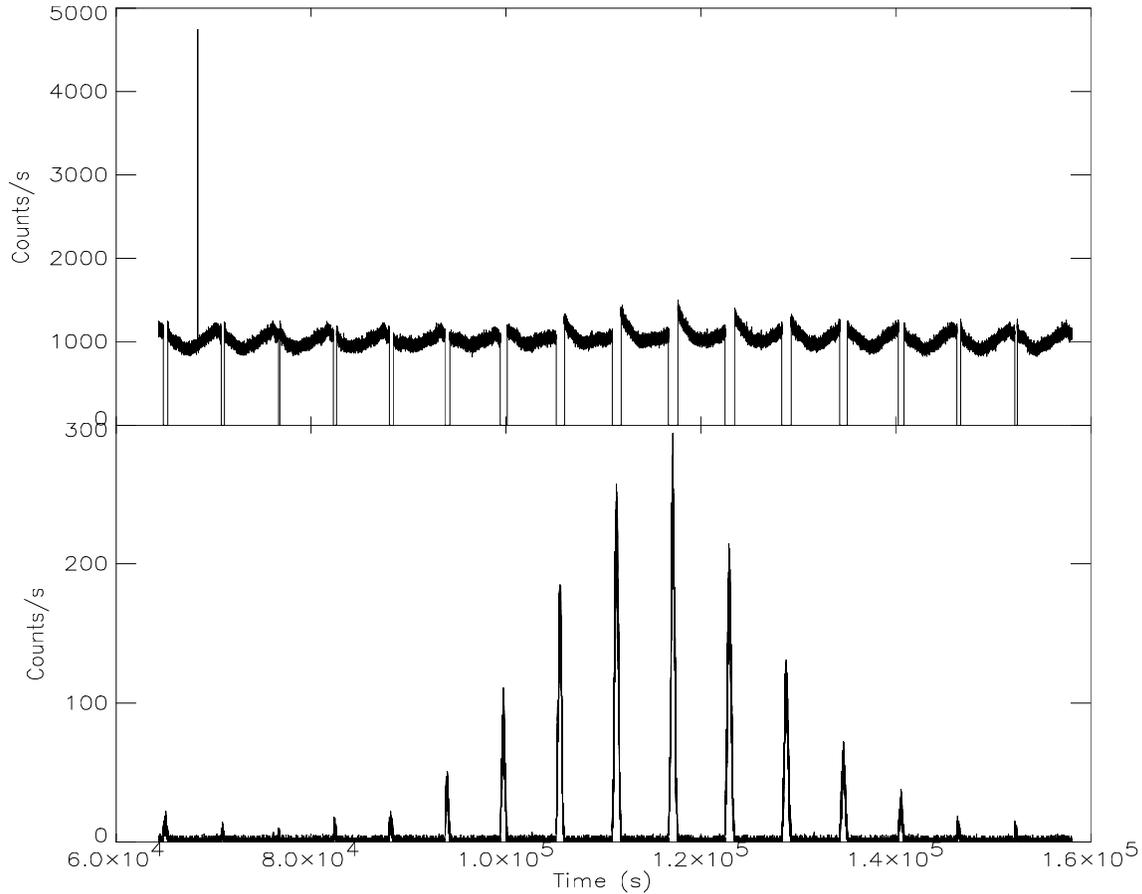

**Figure 3:** Time profile of the Anticoincidence ratemeter LS#1 (top panel) shown on the same time scale as the count rate of the Particle Monitor (bottom panel). The satellite passages close to the South Atlantic Geomagnetic Anomaly are clearly evident from the PM count rate and the corresponding activation effect in the CsI of the shield is evident from the higher background detected as the satellite emerges from the SAGA. The spike on the left is a real gamma-ray burst (GRB970704).

With the above setting of the thresholds we detect about 12 triggers per day, that is roughly 0.8 triggers per orbit. These triggers are uniformly distributed along the orbit and there is no evidence for any "hot spot". Most of them are of course false triggers due to the lack of any particle anticoincidence (the GRBM detectors are themselves an active anticoincidence of the PDS experiment). Actually what we experienced is that the false triggers are due to correlated events in two contiguous shields. The physical origin of this kind of events is the same as the spikes, that is high Z particles that excite metastable atomic states of the crystals, but they are due to the crossing of two shields by the same particle, and therefore in time coincidence. In fact, what we see in our data is that most of the false triggers are composed by fast events (from few to tens of milliseconds) with slightly different time histories in two shields that are always close by. This characteristic, apart from the statistical considerations given above, almost excludes the possibility that these false triggers can be due to independent spikes in time coincidence on two shields. In this case one would expect they should be equally distributed on the four shields and not always from adjacent pairs. Their outlined characteristics make this kind of false triggers indistinguishable from real, short GRBs and we could reject them only by renouncing to the possibility of detecting very short events. On this regard, an interesting point is that we detect a small amount of short events that has similar characteristics to those described above. They are weak, short, on two adjacent shields (a viewing angle effect on a real GRB could also cause this) in which they show a very similar time profile. Among them we think that there can be some real GRB that can only be confirmed through the simultaneous detection by some other experiment.

In the above discussion we did not consider the LIT value, that is the running time interval on which the onboard software computes the mean value of the detector count rate. It was initially set at 128 s, and this could still be an acceptable value thanks to the very good stability of the background. Nevertheless we have faced an unpredicted situation whose origin is not clear and is still under study. In an orbit portion preceding the SAGA by about 10 minutes, sometime we detect an increase in the count rate of all detectors, mainly concentrated in the lower energies (i.e. smaller than 100 keV). The increase is typically negligible, but at times it becomes important, and rarely dramatic. We call this phenomenon "pre-SAGA", and we have tried to investigate its origin. We first have searched for a correlation with solar flares, but we did not find anything significant. Considering that the phenomenon shows up in a fixed portion of the orbit, we expect it could be due to some geomagnetic effect. It should be note that the excess intensity is at its maximum during those orbits with the most distant passage from the SAGA, that are clearly distinguishable thanks to the PM (see Figure 3). In Figure 4 the most dramatic example of the pre-SAGA we have detected is shown. As it is visible the pre-SAGA was exceptionally intense at time around -5000, and it was still intense in the next orbit. We note that during this second orbit, overimposed to the pre-SAGA, we detected the famous GRB970228 (more visible from the expanded views in the two bottom panels). This GRB was triggered by the GRBM thanks to the fact that because of the pre-SAGA effect the LIT was set at 32 s, in order to properly follow these possible variation of the background.

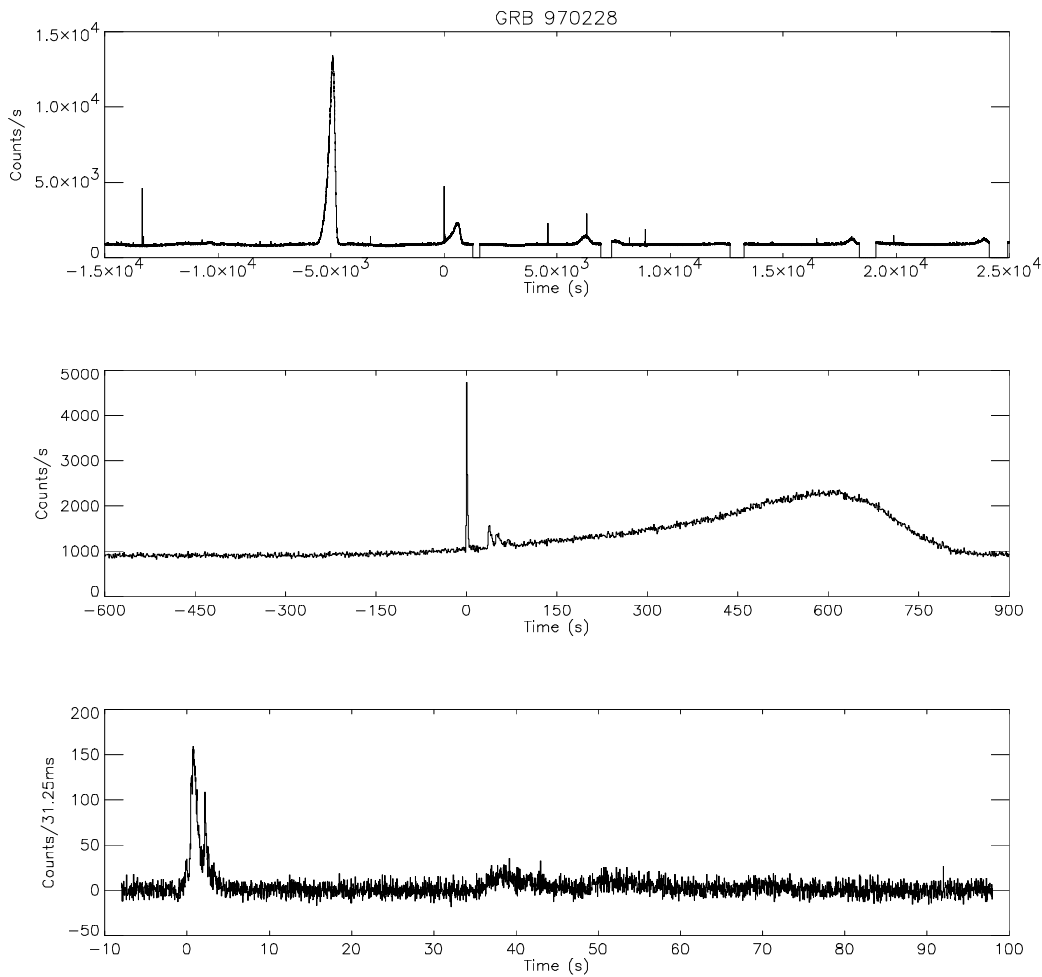

**Figure 4:** GRB970228 as detected in the GRBM #1 ratemeter (T= 0 s, corresponding to the trigger time: 1997, Feb. 28, 02:58:00 UT). The pre-SAGA anomaly is huge in the previous orbit and evident in the orbit of GRB970228 itself. In the middle panel the GRB region is enlarged in order to make GRB970228 visible. In the bottom panel the burst profile (31.25 ms bins) as seen with the high time resolution data, in which the minor structures of the first peak can be clearly identified.

Once the trigger condition is verified in at least two of the four shields within the SIT the high resolution temporal profiles are stored, according to what described in paragraph 2. When we first analysed these data we verified that a bug was present in the onboard software causing that part of the light curve was overwritten and the consequent loss of some temporal data. The bug was fixed and the software patch was loaded onboard on October 1996. The effect of this bug can be seen, as an example, in the time profile of GRB960720 shown in Figure 6, where the overwritten data are substituted with null values. An example of the proper functioning of the high resolution time profiles can instead be seen in the lowest panel of Figure 4 for the case of GRB970228. The high time resolution allows the identification of minor structures in the burst profile, that in the ratemeters are blended to form a single peak.

From Figure 2 it appears that the GRBM ratemeters of the shield #2 has a higher background with respect to the other three. This is due to the fact that the shield #2 is illuminated by the four $Cd^{109}$ radioactive sources (main emission 88 keV, 462 days half life) that are part of the in-flight calibration system of the High Pressure Gas Scintillation Proportional Counter[4] (HPGSPC, see Figure 1), one of the two narrow field, high energy experiments onboard BeppoSAX. The 88 keV line is efficiently detected by the shield #2 in the energy range of the GRBM ratemeters, but due to its higher low energy threshold the LS #2 ratemeter is not affected by the same problem.

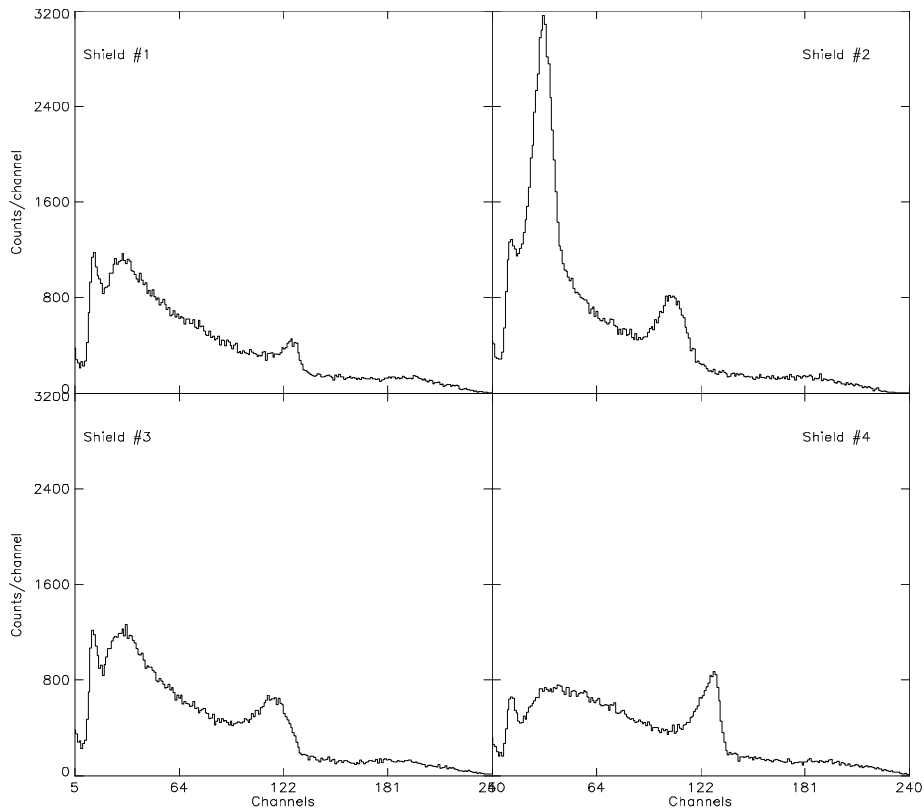

**Figure 5:** In-flight background energy spectra of the four detectors of the GRBM experiment. The 88 keV line due to the HPGSPC calibration radioactive source is clearly visible in the spectrum of shield #2.

On the other hand, LS #4 is influenced by another "local" source of background. This is the linear $Co^{57}$ movable calibration source of the PDS experiment itself, whose parking position is located just above the top of shield #4. The standard emission (122-136 keV) of the source is efficiently shielded by a properly heavy "parking strip", but a secondary line at 692 keV is emitted by the source with a branch of 0.16% (compared to the 96% of the main lines) and due to its high energy it is not efficiently shielded, therefore causing a higher background in the shield #4. All of these characteristics of the background can be seen in the energy spectra presented in Figure 5. In particular the prominent 88 keV line is clearly visible in the GRBM#2 with an energy resolution of about 36% FWHM. In all of the shields is also visible the calibration light source provided by an $Am^{241}$ source embedded in NaI crystal. Actually this kind of calibrator, however, is rather sensitive to temperature variations, and can therefore be used only for comparing situations with similar experiment temperatures.

## 4. GAMMA-RAY BURSTS DETECTION

The BeppoSAX/GRBM has become world wide popular in the last few months thanks to its simultaneous detection of some GRBs with the Wide Field Cameras onboard BeppoSAX itself, and the consequent capability of quickly locating the GRB in the sky with a 3 arcminutes accuracy. However the few GRBs detected also from one of the two WFCs are just a small portion of the GRBs that are continuously detected by the GRBM only. Unfortunately, due to large amount of activity required for the fast detection, analysis and follow-up observations of the GRBM/WFC GRBs, the systematic screening of the GRBM data is still in a large delay and only a minor portion of the available data have been analysed.

In Figure 6 we show few examples of the GRBs detected by the GRBM. Some of them were simultaneously detected by other experiments like the BATSE experiment onboard CGRO or the GRB detector onboard the Ulysses[20] interplanetary probe or the Konus[21] experiment onboard the Wind satellite.

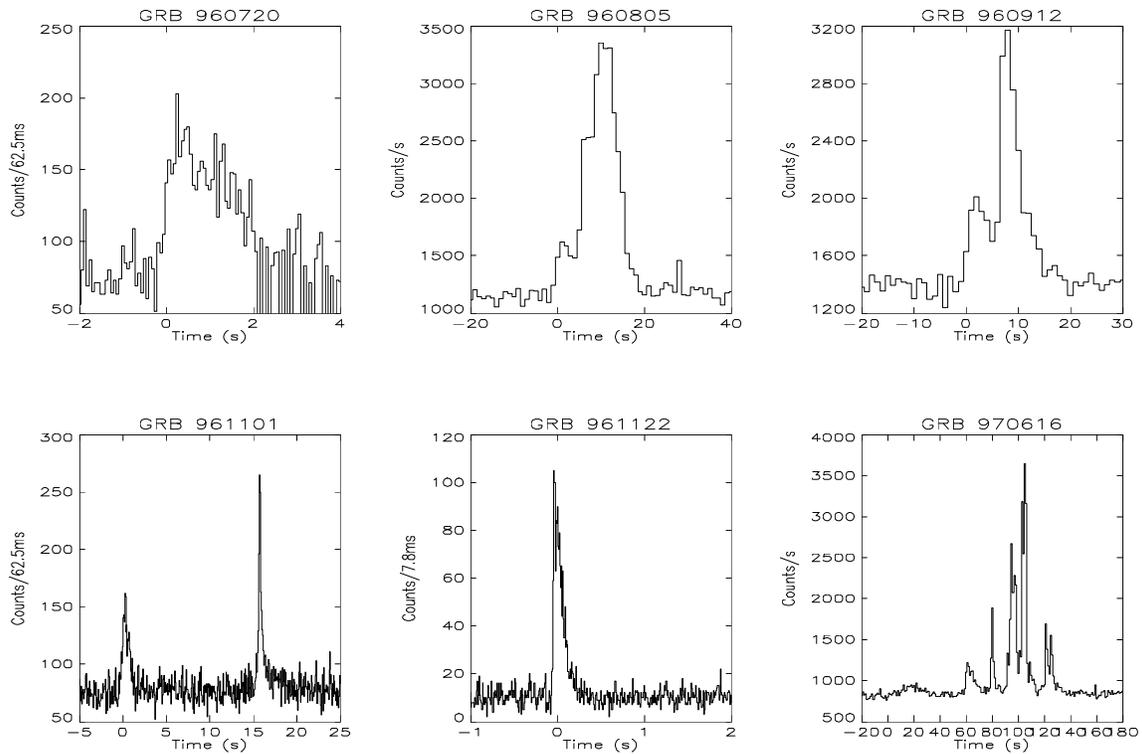

**Figure 6:** Six GRBs detected by the GRBM onboard BeppoSAX. GRB start time is zero on the time axis. From the top left: GRB960720 (20/Jul/1996, 11:36:53 UT, also detected by BATSE) as visible from the high resolution time profiles of shield #1, rebinned at 62.5 ms (data gaps are due to an onboard software bug, later fixed); GRB960805 (05/Aug/1996, 21:55:59 UT, also detected by Ulysses, not detected by BATSE) as visible in the GRBM#3 ratemeter (1 s time resolution); GRB960912 (12/Sep/1996, 13:57:28, also detected by BATSE) from GRBM#1 ratemeter; GRB961101 (01/Nov/1996, 16:07:44 UT, also detected by Konus, not detected by BATSE) at 62.5 ms from shield #2; GRB961122 (22/Nov/1996, 02:17:40 UT, also detected by BATSE) at 7.8 ms from shield #2; GRB970616 (16/Jun/1997, 18:09:48 UT, also detected by BATSE and Ulysses) at 1 s from shield #1.

As can be seen from Figure 6, the BeppoSAX GRBM was able to trigger on several different types of GRBs. As an example, GRB960720 (the first GRB detected[23] also in the WFCs) is a FRED (Fast Rise Exponential Decay) type, and lasted for about 4 seconds; GRB961122 is also FRED type, but it lasts about 500 ms; GRB961101 has a small precursor, and GRB970616 has a very complex structure, even more complicated when seen at higher time resolution.

A systematic analysis has been done in order to investigate the comparative capability of the GRBM with respect to the BATSE experiment, currently the most sensitive GRB experiment. We have searched in our data for the GRBs detected by BATSE in the period from July to November 1996, and publicly distributed through the BACODINE network[24] (now renamed GCN). We found that about 30% of the BATSE events where also detected by the GRBM. A similar investigation has been done again for the period from May to June 1997, and about 43% of the BATSE events were also detected by the GRBM. To understand this results one has to take into account the following facts: the first period under study includes the GRBM Science Verification Phase, during which the instrument parameters were still to be adjusted and the allocated telemetry was enabled to transmit one GRB per orbit; during part of the first period (November) since the soft gamma-ray repeater SGR1806-20 was active the BATSE team adjusted the trigger criterion in order to optimize it for low energy events; both the instruments operate on a low-earth orbit where the Earth continuously obscures about 34% of the sky; the GRBM has a smaller field of view than BATSE. Therefore this test can be considered consistent with the hypothesis that the two instruments have a very similar sensitivity to trigger GRBs. An indirect confirmation of this rough estimate of the relative efficiency between BATSE and GRBM can also be seen in the fact that out of 5 events presently available from GRBM/WFC, only 3 have been detected also by BATSE. Finally, it is worth noticing that two of this unselected sample of 5 are among the weakest GRBs in the *Number* vs. *Peak Flux* distribution of the BATSE events.

## 5. CONTRIBUTION TO THE THIRD INTERPLANETARY NETWORK

Since the beginning of its operation the GRBM has joined the Third Interplanetary Network[25] (IPN). This is an international network of GRB detectors devoted to reconstruct with high precision the GRBs arrival direction by measuring the delay among three detectors placed at interplanetary distances one from each other. At present the only GRB detector operating at a large distance from the Earth is the one onboard Ulysses (currently at a distance of about 2800 light seconds from the Earth). Given the lack of a second GRB detector at interplanetary distances the IPN measurements can presently only be done between Ulysses and one of the low-earth satellite (CGRO, BeppoSAX, Wind). The result is an annulus including the possible arrival directions in the sky. The strip width is dependent on the intensity of the event, its temporal structure and the amplitude of the time delay, and can be as narrow as one arcminute. The BeppoSAX-Ulysses IPN annulus has revealed of particular interest in the case of GRB970228 (not detected by BATSE) because it allowed for decreasing the BeppoSAX/WFC error box of the event[26] and get a more reliable identification of the X-ray[13] and optical[14] counterparts.

## 6. SOURCE DETECTION CAPABILITIES WITH THE EARTH OCCULTATION TECHNIQUE

The GRBM detectors can also be used as an All Sky Monitor. Given their large area they can be able to detect transient hard X-ray sources in the sky at any time. However, since the GRBM experiment has no imaging capabilities, steady sources can only be monitored by using the Earth Occultation Technique, as well as currently done by BATSE[27]. When a bright hard X-ray source is obscured by the Earth during a satellite orbit, its contribution to the detector count rate disappears, and it will be visible again when the source will rise from behind the Earth. Due to the repetitive cycle of the satellite orbit, this situation will occur once per orbit as long as the same satellite pointing is maintained, producing a count rate "step" at any rise/set of the source. One can therefore take advantage of summing several "steps" due to the source in subsequent orbits and therefore reduce the statistical error on the measurement. The occultation profile is smoothed by the Earth atmosphere and in particular it depends on the time needed to the source to cross the atmospheric profile, that is by the inclination of the line of sight to the source with respect to the orbital plane. In Figure 7 we show the occultation step due to the Crab Nebula in one of the four GRBM detectors, obtained with the thresholds configuration given in paragraph 3. Overimposed to the detector time profile is the result of a modelization of the atmospheric absorption. The inferred net Crab count rate is $(40\pm1)$ counts s$^{-1}$.

## 7. CONCLUSIONS AND FUTURE DEVELOPMENTS

In this paper we presented the in-flight performances of the GRBM experiment onboard the BeppoSAX satellite since the beginning of the mission operation. Besides its fundamental role in the recent discovery of X-ray and optical counterparts of some GRBs, the experiment is returning a huge amount of data on GRBs, most of which are still to be analysed. Even if the GRBM is actually a secondary function of a different experiment, its sensitivity turned out to be comparable to that of the dedicated experiment onboard CGRO, BATSE. Furthermore the GRBM has also shown encouraging results for what concern its capability in the source monitoring through the earth occultation method.

An intense activity of calibration[22,28] is being performed by the Hardware Team in order to fully exploit the experiment capabilities. In particular, the completion of the analysis of the on-ground calibration (carried out both at experiment and at satellite level) and a detailed Monte Carlo simulation will soon allow for a GRB direction reconstruction, based on the relative counts detected in the various GRBM detectors. These improvements, together with a major effort in data analysis will lead to the compilation of a first BeppoSAX GRB catalog, that will include both GRBs simultaneously detected by other experiments and GRBs detected by the GRBM alone.

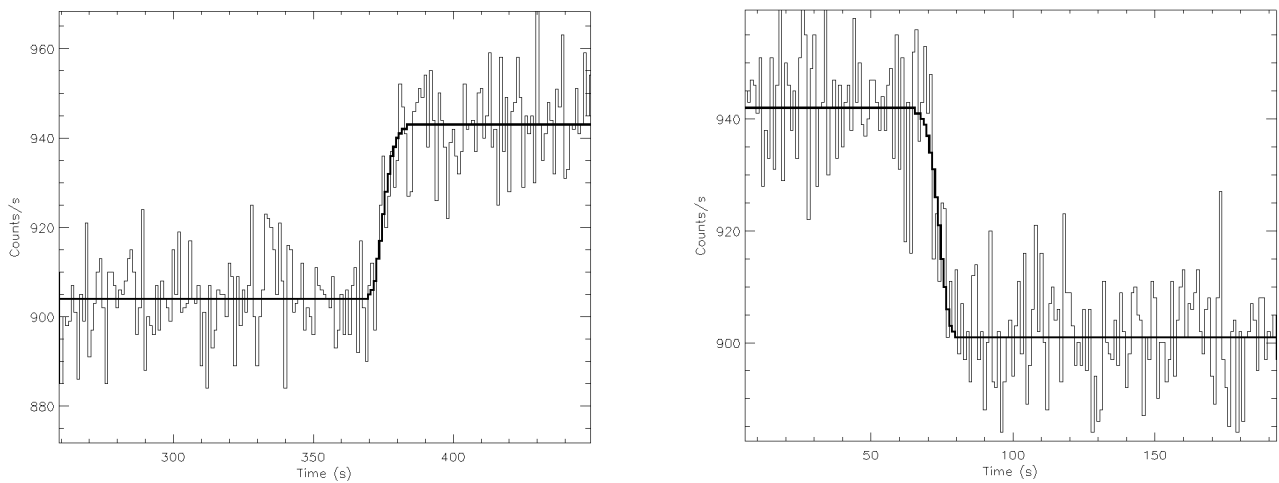

**Figure 7:** Crab Nebula occultation step in the GRBM detector #3, with LLT=3 and ULT=6. Rise and set of the source behind the Earth disk are visible and modelled by an earth atmosphere absorption profile.

## ACKNOWLEDGEMENTS

SAX is a major program of the Italian Space Agency (ASI) with a participation of the Netherlands Agency for Aerospace Programs (NIVR). It was renamed BeppoSAX in honour of Giuseppe (Beppo) Occhialini.

All authors warmly thank the professional teams that worked on this project, and those that are still working, from the Laben (Vimodrone), Alenia Spazio (Torino) and Nuova Telespazio (Roma) companies. A particular thank we would like to devote to P. Attina', G. Cabodi, F. Gilardi and G. Giulianelli (Alenia Spazio), V. Chiaverini, G. Falcetti, F. Monzani and J.M. Poulsen (Laben), L. Salotti, G. Sabatini, C. Aureli, G. Celidonio, G. D'Andreta, G. De Angelis, L. Di Ciolo, G. Gandolfi, R. Ricci, M. Smith, G. Spoliti (Nuova Telespazio) for their specific and enthusiastic contribution to this program.